\documentclass[twoside,twocolumn,english,pra,show pacs]{revtex4}
\usepackage[T1]{fontenc}
\usepackage[latin9]{inputenc}
\usepackage{amsbsy}
\usepackage{amssymb}
\usepackage{graphicx}
\usepackage{esint}

\makeatletter
\@ifundefined{textcolor}{}
{%
 \definecolor{BLACK}{gray}{0}
 \definecolor{WHITE}{gray}{1}
 \definecolor{RED}{rgb}{1,0,0}
 \definecolor{GREEN}{rgb}{0,1,0}
 \definecolor{BLUE}{rgb}{0,0,1}
 \definecolor{CYAN}{cmyk}{1,0,0,0}
 \definecolor{MAGENTA}{cmyk}{0,1,0,0}
 \definecolor{YELLOW}{cmyk}{0,0,1,0}
}

\makeatother

\usepackage{babel}
\begin{document}

\title{Time-of-flight patterns of ultra-cold bosons in optical lattices\\
in various Abelian artificial magnetic field gauges}

\author{T. P. Polak}

\address{Faculty of Physics, Adam Mickiewicz University of Pozna\'{n}, Umultowska
85, 61-614 Pozna\'{n}, Poland}

\author{T. A. Zaleski}

\affiliation{Institute of Low Temperature and Structure Research, Polish Academy
of Sciences, Okólna 2, 50-422 Wroc\l{}aw, Poland}
\begin{abstract}
We calculate the time-of-flight patterns of strongly interacting bosons
confined in two-dimensional square lattice in the presence of an artificial
magnetic field using quantum rotor model that is inherently combined
with the Bogolyubov approach. We consider various geometries of the
magnetic flux, which are expected to be realizable, or have already
been implemented in experimental settings. The flexibility of the
method let us to study cases of the artificial magnetic field being
uniform, staggered or forming a checkerboard configuration. Effects
of additional temporal modulation of the optical potential that results
from application of Raman lasers driving particle transitions between
lattice sites are also included. The presented time-of-flight patterns
may serve as a verification of chosen gauge in experiments, but also
provide important hints on unconventional, non-zero momentum condensates,
or possibility of observing graphene-like physics resulting from occurrence
of Dirac cones in artificial magnetic fields in systems of ultra-cold
bosons in optical lattices. Also, we elucidate on differences between
effects of magnetic field in solids and the artificial magnetic field
in optical lattices, which can be controlled on much higher level
leading to effects not possible in condensed matter physics.
\end{abstract}


\maketitle

\section{Introduction}

The experimental observation of Bose Einstein condensation \cite{anderson}
of trapped atomic gases catalyzed a very large further research activity
in studies of behavior of atoms obeying Bose-Einstein statistics.
This led to loading of ultra-cold bosonic atoms into optical lattices,
which offer a clean setting for quantitative and highly precise investigations
of quantum phase transitions in the strongly interacting atomic systems
\cite{greiner}. The behavior of the atoms bears resemblance to the
physics of strongly-correlated electronic systems like high-T$_{c}$
superconductors, which are described by similar microscopic Hamiltonians
(Hubbard model) \cite{jaksch1}. Although the particles that are loaded
to optical lattices are electrically neutral, it is possible to impose
additional external potential, which forces them to behave exactly
like charged particles interacting with an external magnetic field.
In the simplest case the potential can result from rotation, following
from formal equivalence between the Lorentz force and the Coriolis
force \cite{leggett,cooper}. However, more control is obtained using
additional photon-assisted tunneling to coherently transfer atoms
from one internal state to another. This induces a non-vanishing phase
of particles moving along a closed path, which simulates magnetic
flux through the lattice \cite{jaksch,cooper,gerbier,kolovsky}. Such
techniques are under very active investigation. Since, quantum optics
technology provides unprecedented degree of manipulation of structure
of such imposed magnetic flux, it allows for obtaining very strong
magnetic fields for neutral atoms: both Abelian \cite{lin,aidelsburger}
and non-Abelian \cite{osterloh,hauke}. A long-term goal of these
experiments is to achieve the quantum Hall regime, in which very high
value of effective flux opens an avenue to study the effects that
are not achievable in the conventional solid state physics \cite{harper,hofstadter}. 

The dynamic properties of the trapped ultra-cold atoms can be investigated
in time-of-flight (TOF) experiments, in which the trapping potential
is suddenly switched off. As the atoms are no longer being localized
spatially, their scattering ratio decreases and the momentum distribution
becomes temporarily frozen, with gravity being the only force acting
on the atomic cloud. The infra-red absorption images taken after arbitrary
expansion time show locations of the atoms, which are directly related
to the the distribution of the momenta $n\left(\boldsymbol{k}\right)$
in the system before the potential was switched off. In the superfluid
state (SF) the TOF images exhibit characteristic sharp maxima related
to long-range phase coherence of the condensate. While the phase fluctuations
are being increased by stronger interactions between atoms, the sharp
features disappear and $n\mbox{\ensuremath{\left(\mathbf{k}\right)}}$
becomes a wide maximum indicating the presence of the Mott insulating
(MI) state \cite{greiner}. 

Theoretical challenge in describing the TOF patterns results from
the dynamic nature of the problem: determination of the momentum distribution
$n\left(\mathbf{k}\right)$ requires precise knowledge of spatial
correlations between atoms, namely atom-atom correlation function.
This precludes use of methods based on mean-field approximation. In
the present paper we apply a recently proposed combination of quantum
rotor (QR) approach and Bogolyubov method, which has been successfully
applied to investigate correlations in systems of cold atoms in optical
lattices (e.g. time-of-flight patterns \cite{zaleski}, spectral functions
\cite{zaleski1}). It is also a natural extension of the QR model
that was used to describe the phase diagram, also in the presence
of the artificial magnetic field \cite{polak1,sinha,polak}. The QR
approach has been verified \cite{polak2,polakzaleski1} using other
methods, like Monte Carlo numerical calculations \cite{capogrosso-sansone}
and diagrammatic perturbation theory \cite{teichmann}. The QR phase
diagrams were also analyzed in the context of an analytical works:
mean-field theory \cite{oktel} and Pad\'{e} analysis \cite{niemeyer}.

It is our goal to calculate the TOF patterns for various gauges of
the magnetic field that are expected to be realizable or have already
been implemented in the experimental settings to investigate the dynamics
of the condensate and the phase transition to the localized state.
The remainder of the paper is as follows: in. Sec II we introduce
the model Hamiltonian relevant to strongly correlated bosons confined
in two-dimensional square lattice. Then we apply the synthetic magnetic
field\textbf{,} which modifies the hopping term in the Hamiltonian
and add additional temporal modulation of the optical potential. The
main points of our approach that lead to calculation of the atom-atom
correlation function are summarized in Sec. III. Furthermore, we present
the time-of-flight patterns in the following Section. Our results
are summarized in Sec. V, while the dispersion relations and the resulting
lattice densities of states used in calculations are presented in
Appendix.

\section{Model Hamiltonian}

The essential physics of bosons in optical lattice can be captured
using the single-band Bose-Hubbard model. In this description, the
particles move within a tight-binding scheme and interact only through
on-site repulsion resulting from interatomic collisions (since the
atoms are neutral). The Hamiltonian is given by, 
\begin{equation}
\mathcal{H}=-\sum_{\left\langle i,j\right\rangle }t_{ij}\left(b_{i}^{\dagger}b_{j}+b_{j}^{\dagger}b_{i}\right)+\frac{U}{2}\sum_{i}n_{i}\left(n_{i}-1\right)-\mu\sum_{i}n_{i},\label{hamiltonian}
\end{equation}
where $b_{i}\left(b_{i}^{\dagger}\right)$ is the boson destruction
(creation) operator at a site $i$, $n_{i}=b_{i}^{\dagger}b_{i}$
is the density operator, $U>0$ is the on-site repulsion and $\mu$
is the chemical potential, which controls the number of bosons. Here,
$\left\langle i,j\right\rangle $ denotes summation over the nearest-neighbor
sites. Finally, $t_{ij}$ is the hopping matrix element, which is
non-zero only for the nearest neighbors and equal to $t$. For any
given lattice geometry and depth, both $t$ and $U$ can be calculated
directly by finding the respective Wannier function basis \cite{blackie}.
Introduction of the synthetic magnetic field $\mathbf{B}$ (potential
that acts on neutral particles in the same fashion as the magnetic
field acts on charges, e.g. rotation of the system, laser stirring,
selective driving of hopping with Raman lasers) leads to introduction
of the Peierls phase factor:
\begin{equation}
\exp\left(\frac{2\pi i}{\Phi_{0}}\int_{\mathbf{r}_{i}}^{\mathbf{r}_{j}}\mathbf{A}\cdot d\mathbf{l}\right),
\end{equation}
which is a consequence of the gauge invariance of the Schr\"{o}dinger
equation, where $\mathbf{B}=\nabla\times\mathbf{A}\left(\boldsymbol{r}\right)$
and $\Phi_{0}=h/e$ is the flux quantum, with $\mathbf{A}\left(\boldsymbol{r}\right)$
being the vector potential. This leads to modification of the hopping
term:
\begin{eqnarray}
-\sum_{\left\langle i,j\right\rangle }t_{ij}\left(b_{i}^{\dagger}b_{j}+b_{i}b_{j}^{\dagger}\right) & \rightarrow & -\sum_{\left\langle i,j\right\rangle }t_{ij}\left(b_{i}^{\dagger}b_{j}e^{\frac{2\pi i}{\Phi_{0}}\int_{\mathbf{r}_{i}}^{\mathbf{r}_{j}}\mathbf{A}\cdot d\mathbf{l}}\right.\nonumber \\
 &  & \left.+b_{i}b_{j}^{\dagger}e^{-\frac{2\pi i}{\Phi_{0}}\int_{\mathbf{r}_{i}}^{\mathbf{r}_{j}}\mathbf{A}\cdot d\mathbf{l}}\right).
\end{eqnarray}
that as a result, instead of being a real value, becomes a complex
number:
\begin{equation}
t_{ij}\rightarrow t_{ij}^{'}\equiv t_{ij}e^{\frac{2\pi i}{\Phi_{0}}\int_{\mathbf{r}_{i}}^{\mathbf{r}_{j}}\mathbf{A}\cdot d\mathbf{l}}.\label{eq:complex_tij}
\end{equation}
Furthermore, particles hopping along closed loops of the lattice cell
(area $a$) gain an additional phase $\phi_{\mathrm{}}\equiv2\pi f$
imposed by the external uniform synthetic magnetic field potential,
where $f=aBe/2\pi\hbar$. We also permit for additional spatial modification
of the on-site potential $\left(-1\right)^{i}\Delta$, which allows
us to describe the effect of temporal modulation by the photon-assisted
tunneling that are used to drive the phase change in some experiments.
The range of actual gauges (shapes of the vector potential $\mathbf{A}$)
applied to the system, which can be realized experimentally, is very
wide (see, Sec. IV). The change of the hopping parameter in Eq. (\ref{eq:complex_tij})
also modifies the band structure, which becomes very complex. Complicated
multi-band dispersion relations provide difficulties in calculating
analytical formulas in the uniform case for the lattice density of
states (DOS) limiting availability to a few selected values of $f$
\cite{polak}.

\section{Correlation functions}

In optical lattices, the phase transition between superfluid and Mott
insulator states occurs in the regime of intermediate to strong interactions
($U\gg t$). As a result, a theory that goes beyond standard Bogoliubov
approximation is required. To this end, we calculate the one-particle
correlation function that is necessary to predict the time-of-flight
patterns using the quantum rotor approach (see, Ref. \cite{polak1})
combined with the Bogolyubov method that has been recently proposed
and succesfully applied to systems of bosons in optical lattices \cite{zaleski}.
This scenario provides a picture of quasiparticles and energy excitations
in the strong interaction limit, where the transition between the
superfluid and the Mott state is driven by phase fluctuations. The
approach is based on separation of the problem into the amplitude
of the Bose field and the fluctuating phase that was absent in the
original Bogoliubov problem. As a results, one arrives at a formalism,
where the one-particle correlation functions are treated self-consistently
and permit us to investigate a whole range of phenomena described
by the Bose-Hubbard Hamiltonian. Furthermore, the phase fluctuations
are described within the quantum spherical model \cite{vojta}, which
goes beyond mean-field approximation including both quantum and spatial
correlations. Although, the approach easily allows for non-zero temperatures,
in the following we restrict ourselves to the description of the ground
state of the system ($T=0$). As the details of calculations have
been extensively presented in Ref. \cite{zaleski}, we only summarize
the main steps of the approach here. We start by introducing the functional
integral representation of the model in Eq. (\ref{hamiltonian}) in
terms of the the complex fields $a_{i}\left(\tau\right)$, which leads
to the partition function:
\begin{equation}
\mathcal{Z}=\int\left[\mathcal{D}\bar{a}\mathcal{D}a\right]e^{-\mathcal{S}\left[\bar{a},a\right]}
\end{equation}
with the action $\mathcal{S}$ given by 
\begin{equation}
\mathcal{S}[\bar{a},a]=\sum_{i}\int_{0}^{\beta}d\tau\left[\bar{a}_{i}\left(\tau\right)\frac{\partial}{\partial\tau}a_{i}\left(\tau\right)+\mathcal{H\left(\tau\right)}\right],\label{action}
\end{equation}
 where $\beta=1/k_{\mathrm{B}}T$ and $T$ being temperature. Next,
we perform the local gauge transformation to the new bosonic variables
\begin{equation}
a_{i}\left(\tau\right)=b_{i}\left(\tau\right)\exp\left[i\varphi_{i}\left(\tau\right)\right].\label{eq:var_transform}
\end{equation}
It allows to extract phase variable $\varphi_{i}\left(\tau\right)$,
which ordering naturally describes the superfluid -- Mott insulator
transition, and the amplitude $b_{i}\left(\tau\right)$ that is related
to the superfluid density. As a result, the the partition function
becomes:
\begin{equation}
\mathcal{Z}=\int\left[\mathcal{D}\bar{b}\mathcal{D}b\right]\left[\mathcal{D}\varphi\right]e^{-\mathcal{S}\left[\bar{b},b,\varphi\right]},\label{eq:part_fun_bfi}
\end{equation}
with the action $\mathcal{S}\left[\overline{b},b,\varphi\right]\equiv S\left[\overline{a},a\right]$.
The statistical sum in Eq. (\ref{eq:part_fun_bfi}) can be integrated
over the phase or amplitude variables leading to phase-only of amplitude-only
actions:
\begin{eqnarray}
\mathcal{S}_{\varphi}\left[\varphi\right] & = & -\ln\int\left[\mathcal{D}\varphi\right]e^{-\mathcal{S}\left[\varphi\right]},\nonumber \\
\mathcal{S}_{b}\left[\overline{b},b\right] & = & -\ln\int\left[\mathcal{D}\bar{b}\mathcal{D}b\right]e^{-\mathcal{S}\left[\bar{b},b\right]},\label{eq:SfiSbb}
\end{eqnarray}
to obtain:
\begin{equation}
Z=\int\left[\mathcal{D}\varphi\right]e^{-\mathcal{S}_{\varphi}\left[\varphi\right]}=\int\left[\mathcal{D}\bar{b}\mathcal{D}b\right]e^{-\mathcal{S}_{b}\left[\bar{b},b\right]}.
\end{equation}
The main point of the approach is the calculation of the action $\mathcal{S_{\varphi}\left[\varphi\right]}$
in Eq. (\ref{eq:SfiSbb}), which describes the phase-only model with
amplitudes integrated out. It is subsequently mapped onto the quantum
spherical model, which can be solved analytically. 

As a result of the variable transformation in Eq. (\ref{eq:var_transform}),
the superfluid order parameter, which non-vanishing value signals
a macroscopic quantum phase coherence (identified as the superfluid
state), factorizes: 
\begin{equation}
\Psi_{B}\equiv\left\langle a_{i}\left(\tau\right)\right\rangle _{a}=\left\langle b_{i}\left(\tau\right)\right\rangle _{b}\left\langle \exp\left[i\varphi_{i}\left(\tau\right)\right]\right\rangle _{\varphi}.\label{eq:ordparfact}
\end{equation}
This reflects the fact that all atoms in the condensate form a coherent
matter wave having the same phase. The averages in Eq. (\ref{eq:ordparfact})
are defined as:
\begin{equation}
\left\langle \dots\right\rangle _{x}=\frac{\int\left[\mathcal{D}x\right]\dots e^{-\mathcal{S}_{x}\left[x\right]}}{\int\left[\mathcal{D}x\right]e^{-\mathcal{S}_{x}\left[x\right]}}
\end{equation}
for $x=a,\, b,\,\varphi$ and the respective actions: $\mathcal{S}\left[\overline{a},a\right]$,
$\mathcal{S}\left[\overline{b},b\right]$ or $\mathcal{S}_{\varphi}\left[\varphi\right]$.
Furthermore, we parametrize the boson fields 
\begin{equation}
b_{i}\left(\tau\right)=b_{0}+b_{i}^{'}\left(\tau\right),\label{parametrization}
\end{equation}
where $b_{0}=\sqrt{N_{0}}$ is the Bose condensate macroscopic occupation
and $b_{i}'\left(\tau\right)$ is the amplitude fluctuation around
the mean value $b_{0}$. As a result, the superfluid order parameter
becomes:
\begin{equation}
\Psi_{B}=b_{0}m_{0},
\end{equation}
where $m_{0}$ is phase order parameter:
\begin{equation}
m_{0}=\left\langle \exp\left[i\varphi_{i}\left(\tau\right)\right]\right\rangle .
\end{equation}
The atom-atom correlation function 
\begin{equation}
C_{ij}\left(\tau\right)=\left\langle a_{i}\left(\tau\right)\overline{a}_{j}\left(\tau\right)\right\rangle _{x}\label{correlation function}
\end{equation}
also factorizes due to the variable transformation in Eq. (\ref{eq:var_transform})
becoming: 
\begin{equation}
C_{ij}\left(\tau\right)=\left\langle b_{i}\bar{b}_{j}\right\rangle _{b}\left\langle \exp\left[\varphi_{i}\left(\tau\right)-\varphi_{j}\left(\tau\right)\right]\right\rangle _{\varphi}\label{eq:correlation_function_bfi}
\end{equation}
with the averages that can be calculated analytically for any lattice,
for which the dispersion relation $t_{\mathbf{k}}$ (Fourier transform
of the hopping $t_{ij}$) is known \cite{zaleski}. The momentum distribution
of the atoms in optical lattice is then a Fourier transform of the
correlation function:
\begin{equation}
n\left(\mathbf{k}\right)=\int_{0}^{\beta}d\tau\sum_{\boldsymbol{R}=\left|\boldsymbol{r}_{i}-\mathbf{r}_{j}\right|}C_{ij}\left(\tau\right)e^{i\mathbf{kR}}.
\end{equation}
This leads to the density of particles in the time-of-flight experiments
\cite{kato,zaleski}: 
\begin{equation}
n\left(\mathbf{r}\right)=\left(\frac{m}{\hbar t_{e}}\right)^{3}\left|W\left(\mathbf{k}=\frac{m}{\hbar t_{e}}\mathbf{r}\right)\right|^{2}n\left(\mathbf{k}=\frac{m}{\hbar t_{e}}\mathbf{r}\right),\label{density distribution}
\end{equation}
where $\left|W\left(\mathbf{k}\right)\right|$ is the envelope of
the Fourier transform of the Wannier function for the chosen optical
lattice and $t_{e}$ is the expansion time. It should be pointed out
that the envelope $\left|W\left(\mathbf{k}\right)\right|$ can in
principle depend not only on the optical lattice potential, but also
on the presence of the artificial magnetic field \cite{powell}. However,
since its calculation goes beyond the scope of the present work, we
use the standard form:
\begin{equation}
\left|W\left(\frac{m\mathbf{r}^{2}}{\hbar t_{e}}\right)\right|^{2}\approx\frac{1}{\pi^{3/2}w_{t}}\exp\left(-\frac{\mathbf{r}^{2}}{w_{t}^{2}}\right),\label{eq:wannier}
\end{equation}
where $w_{t}=\hbar t_{e}/mw_{0}$ with $w_{0}$ being the size of
the on-site Wannier function \cite{zaleski}. This choice can be justified
by comparing the resulting TOF patterns with experimental ones \cite{aidelsburger}
and observing the conformity of the particle density decays as a function
of $\mathbf{r}$ in both cases (see, Sec. \ref{sub:Uniaxially-staggered-flux}).

\section{Experimental Gauges of the Synthetic Magnetic Field\label{sec:Experimental-Gauges-of}}

The momentum distribution is an important observable since it allows
to identify whether the atoms in the optical lattice are in the superfluid
or Mott insulating state. Although, the existence of sharp peaks in
the time-of-flight images that used to be unequivocally associated
with the emergence of the superfluidity is not believed to be sufficient
criterion \cite{kato}, recent analysis have shown that the momentum
distribution can be used to make pretty accurate estimations about
location of the critical regime \cite{pollet}. 

In the following, we calculate the time-of-flight patterns resulting
from Eq. (\ref{density distribution}) for various gauges of the artificial
magnetic field. We start from determining the dispersion relation
$t_{\mathbf{k}}$ for the chosen gauge, which allows us to obtain
the atom-atom correlation function in Eq. (\ref{eq:correlation_function_bfi})
by using the procedure described in details in Ref. \cite{zaleski}. 

We present all the results along similar scheme: first we plot $t_{\mathbf{k}}$
for the chosen gauge and then the calculated time-of-flight patterns
in the superfluid state, near the SF-MI phase transition and in the
Mott insulator. We measure the interaction strength between atoms
using the experimental quantities $V_{0}$ and $E_{R}$ instead of
$t$ and $U$, where $V_{0}$ is the optical potential depth and $E_{R}$
-- the recoil energy. The relation of $V_{0}/E_{R}$ to $t/U$ is
presented in Ref. \cite{zaleski}. One should also note that the minimum
of kinetic energy in the Hamiltonian in Eq. (\ref{hamiltonian}) corresponds
to the maximum of $t_{\mathbf{k}}$ due to the minus sign in the hopping
term. 

\begin{widetext}

\begin{figure}
\includegraphics[scale=0.43]{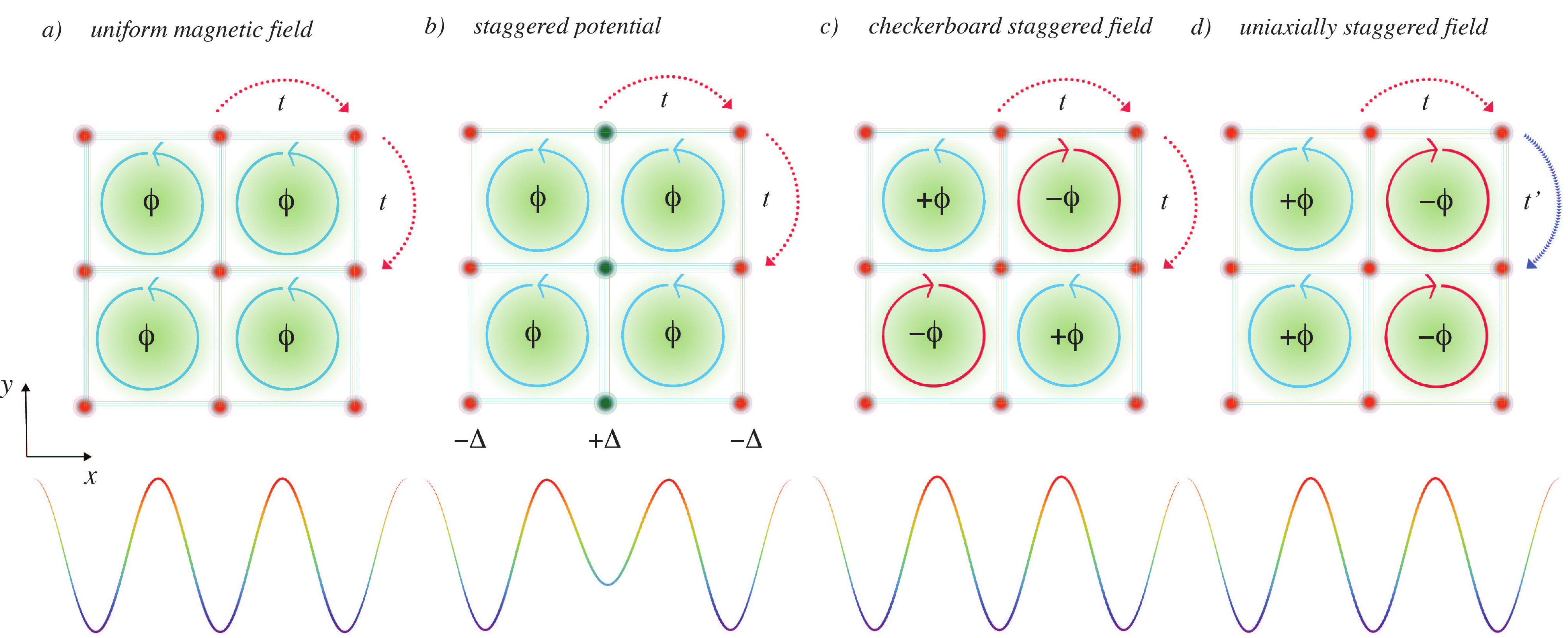}\caption{(Color online) Geometry of the artificial magnetic field in various
gauges resulting in different flux configuration: a) uniform, b) uniform
with additional staggered potential, c) staggered flux with checkerboard
arrangement, d) uniaxially staggered flux. Value of the flux per plaquette
being the phase acquired by a particle traveling around an elementary
cell is $\phi$. \label{fig:1} Additional staggering on-site potential
is denoted by $\pm\Delta$.}
\end{figure}

\end{widetext}

\subsection{Uniform magnetic field}

We start with the uniform artificial magnetic field, which acts on
atoms in optical lattices in identical way as a homogeneous magnetic
fields acts on electrons in solids (see, Fig. \ref{fig:1}a). Every
elementary cell of the lattice is pierced by a fraction $f$ of the
elementary flux, which leads to additional phase $\phi_{\mathrm{}}$
acquired by a particle moving around the cell equal to $\phi\equiv2\pi f$.
Such a configuration of the artificial magnetic field can be realized
using various gauges, e.g. Landau $\mathbf{A}=B\left(0,y,0\right)$,
or symmetric $\mathbf{A}=\frac{B}{2}\left(-y,x,0\right)$. This results
in increase of the elementary cell, since translational symmetry is
locally broken for non-integer values of $f$. If $f$ is a rational,
being equal to $f=p/q$, the cell enlarges $q$-fold, while the Brillouin
zones shrinks by the same factor (leading to so-called magnetic Brillouin
zone). The quasiparticle spectrum has a complicated multi-band structure
known as the Hofstadter butterfly \cite{hofstadter} (the denominator
$q$ determines the number of sub-bands) and can be generated using
Harper's equation \cite{harper}. Although, the general solution is
unknown, for special values of $f$ equal to $f=1/2,1/3,1/4,1/6,1/8,3/8$
both dispersion relation and lattice density of states have been analytically
calculated \cite{polakzaleski1}. We see, that the denominator of
the expression describing the magnetic field $f=p/q$ determines the
number of bands.
\begin{figure}
\includegraphics[scale=0.6]{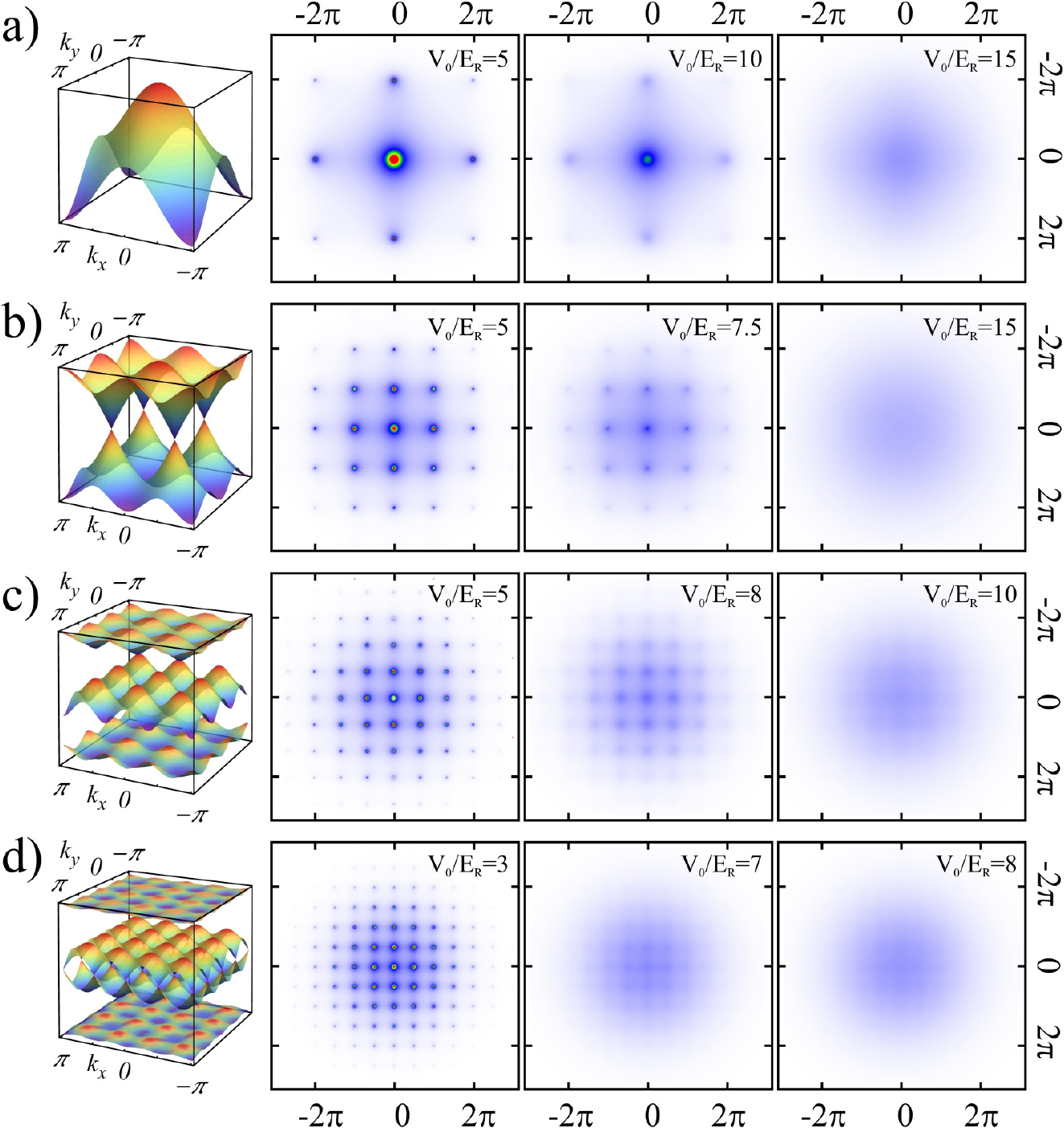}\caption{(Color online) Dispersion relation and time-of-flight patterns for
atoms in optical lattice without (a) and under uniform artificial
magnetic field: b) $f=1/2$, c) $f=1/3$, d) $f=1/4$. The TOF images
depict superfluid state (the first column), system close to the phase
transition (the second column) and the Mott insulator (the last column).}

\label{fig:2}
\end{figure}

The dispersion relations for $f=0,\,1/2,\,1/3$ and $1/4$ have been
presented in Fig. \ref{fig:2} along with the resulting TOF patterns.
In the superfluid state, the spectrum has sharp coherence peaks, which
slowly fade away, when the system is driven towards the phase transition.
Finally, in the Mott insulator, the TOF patterns become a wide, feature-less
maximum with no signatures of the phase coherence. The most notable
influence of the synthetic magnetic field is a change in periodicity
leading to shrinking of the Brillouin zone and accompanying dense
packing of the coherence peaks in the SF state. As a result, small
fluxes could be difficult to detect. In the typical TOF experiment
the $2\hbar k$ of the Brillouin zone corresponds to the $50$ pixels
on the charge-coupled-device (CCD) camera. The smallest detectable
separation between two momentum peaks is around $10$ pixels which
matches the $1/5$ of the Brillouin zone, therefore the fluxes below
$\phi_{\mathrm{}}<2\pi/5$ ($f=1/5$) will be hardly recognizable
from the experimental data. Moreover, the key role is the proper preparation
of the ground state, in which the coherence over large area of the
real space is obtained to avoid further peaks broadening \cite{aidelsburger1}.
In the theoretical calculations the resolution of the TOF diagrams
can be in principle arbitrarily high and two peaks will become indistinguishable
when the distance between them becomes of the order of the full width
at half maximum of the $n\left(\boldsymbol{k}\right)$ peaks, which
in this case is about 3.5\% of the first Brillouin zone width.

\subsubsection{$f=1/2$ $\left(\phi_{\mathrm{}}=\pi\right)$\label{sub:uniform_1p2}}

\begin{figure}
\includegraphics[scale=0.6]{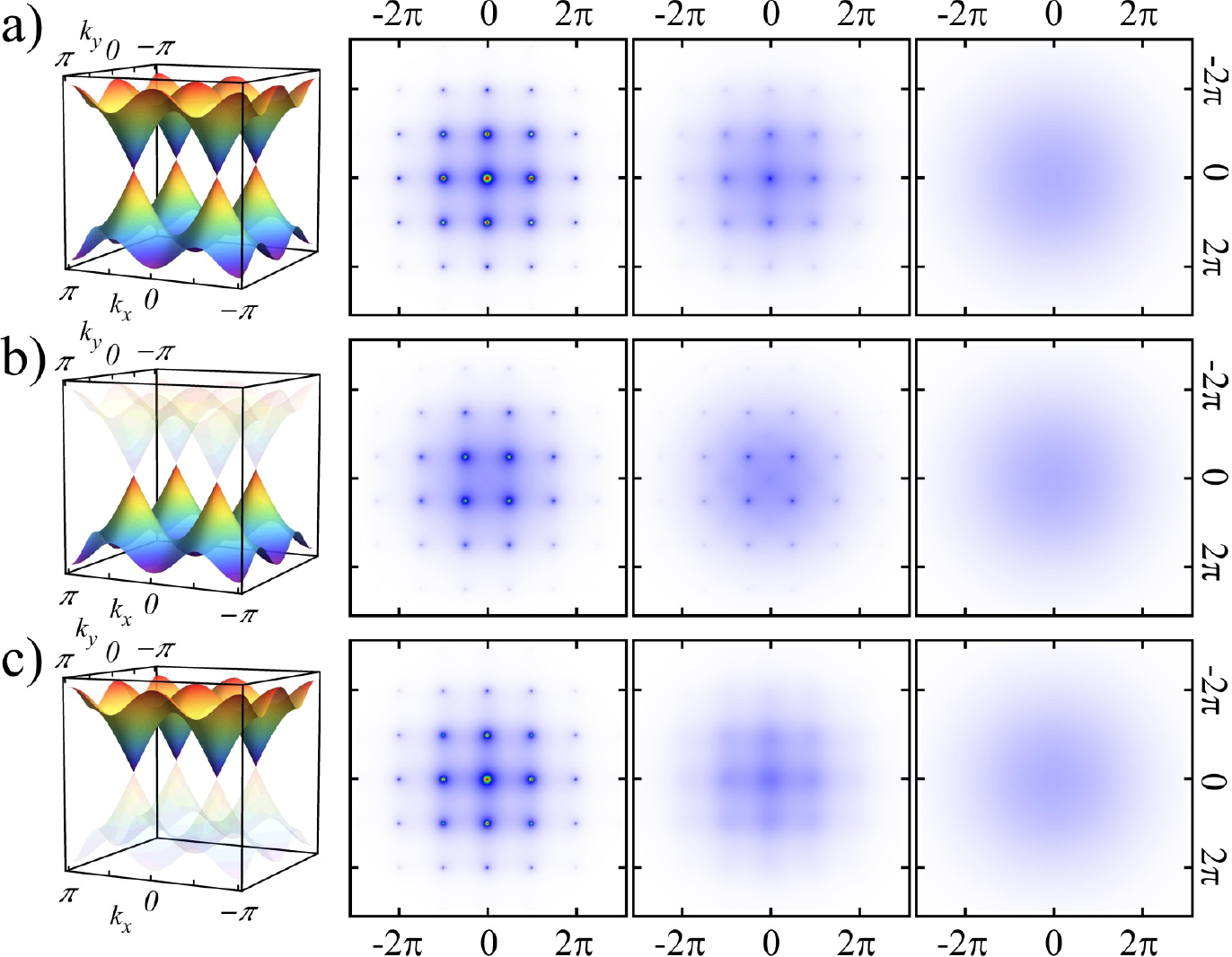}

\caption{(Color online) Dispersion relation and time-of-flight patterns for
atoms in optical lattice under uniform artificial magnetic field $f=1/2$:
a) both sub-bands filled, b) higher sub-band filled, c) lower sub-band
filled. The TOF images depict superfluid state (the first column,
$V_{0}/E_{R}=5$), system close to the phase transition (the second
column, $V_{0}/E_{R}=7.5$) and the Mott insulator (the last column,
$V_{0}/E_{R}=15$).}
\label{fig:3}
\end{figure}

Since, the single-particle spectrum is symmetric around $f=1/2$:
$t_{\mathbf{k}}(f)\equiv t_{\mathbf{k}}(1-f)$, the strongest possible
uniform artificial magnetic field that can be achieved is $f=1/2$,
which results in flux $\phi_{\mathrm{}}=\pi$ per plaquette. In this
case $t_{\mathbf{k}}$ has two sub-bands, which meet at $t_{\mathbf{k}_{D}}=0$
forming Dirac cones (the spectrum is linear near $\mathbf{k}_{D}$
and rotationally symmetric: $t_{\mathbf{k}}\sim\left|\mathbf{k}-\mathbf{k}_{D}\right|$).
This leads to potential possibility of observing graphene-like physics
in optical lattices. The time-of-flight patterns for $f=1/2$ are
presented in Fig. \ref{fig:3}. In the case when bosons are free to
occupy any of the sub-bands (see, Fig. \ref{fig:3}a), the TOF images
show coherence peaks in the superfluid state at the momenta, for which
the kinetic energy assumes minimal values ($\mathbf{k}=\left\{ n\pi,m\pi\right\} ,$where
$n,m$ are integers, which are basically $\mathbf{k}=0$ point repeated
by periodicity of the reciprocal lattice). The $\mathbf{k}_{D}$ points
corresponding to intersections of the bands do not show in the TOF
patterns. However, if \emph{all} particles were occupying \emph{only}
the upper band, the kinetic energy minima would appear at $\mathbf{k}_{D}$,
which has been presented in Fig. \ref{fig:3}b. The resulting superfluid
state exhibits non-zero momentum, with $\mathbf{k}=0$ component totally
removed. On the other hand, in case of populating the lower band only
(see, Fig. \ref{fig:3}c), the resulting picture does not differ much
from the scenario, when occupation of both bands is allowed (in Fig.
\ref{fig:3}a), however some slight differences are noticeable. It
results from the fact that although we are investigating the ground
state of the system, not all bosons occupy the lowest energy state.
Since the particles are interacting, only a fraction of them contributes
to the condensate (thus occupies the lowest energy state), while the
rest can be driven to higher energy states by the quantum fluctuations,
which are present even in zero temperature. This also leads to a conclusion
that in systems of bosons the properties of the superfluid state are
determined by the points in the $k$-space around the minima of the
kinetic energy. However, in the vicinity of the SF-MI phase transition
the sharp maxima are gone, and the TOF images depict momentum distribution
of the incoherent particles. Surprisingly, they contain weak maxima
around $\mathbf{k}=0$ points regardless of location of the superfluid
peaks. 

In order to observe effects resulting from existence of the Dirac
cones in the excitation spectra, population of the respective bands
has to be engineered. It can be experimentally realized using projection
of the condensates onto a desired Bloch state \cite{denschlag}: the
system of bosons, which initially is in the superfluid phase, is released
from the trap and expands freely for a short period of time. Then,
a moving optical lattice is introduced, which is created by laser
beams with additional acousto-optic modulators that allow for shifting
positions of the lattice minima. As a result, depending on the modulation,
the BEC can be loaded to a lattice state with an arbitrary and well-defined
quasi-momentum. Using this approach, it was possible to access different
energy bands of the $^{87}\mathrm{Rb}$ atoms allowing a high precision
studies of the lensing effect on a Bose-Einstein condensate \cite{fallani}.

It should be also noted that the appearance of the Dirac intersections
in the Hofstadter spectra occurs for magnetic fields $f=p/q$, for
which $q$ is even and is never observed for odd values Fig. \ref{fig:2}. 

\subsection{Uniaxial staggered potential}

\begin{figure}
\includegraphics[scale=0.6]{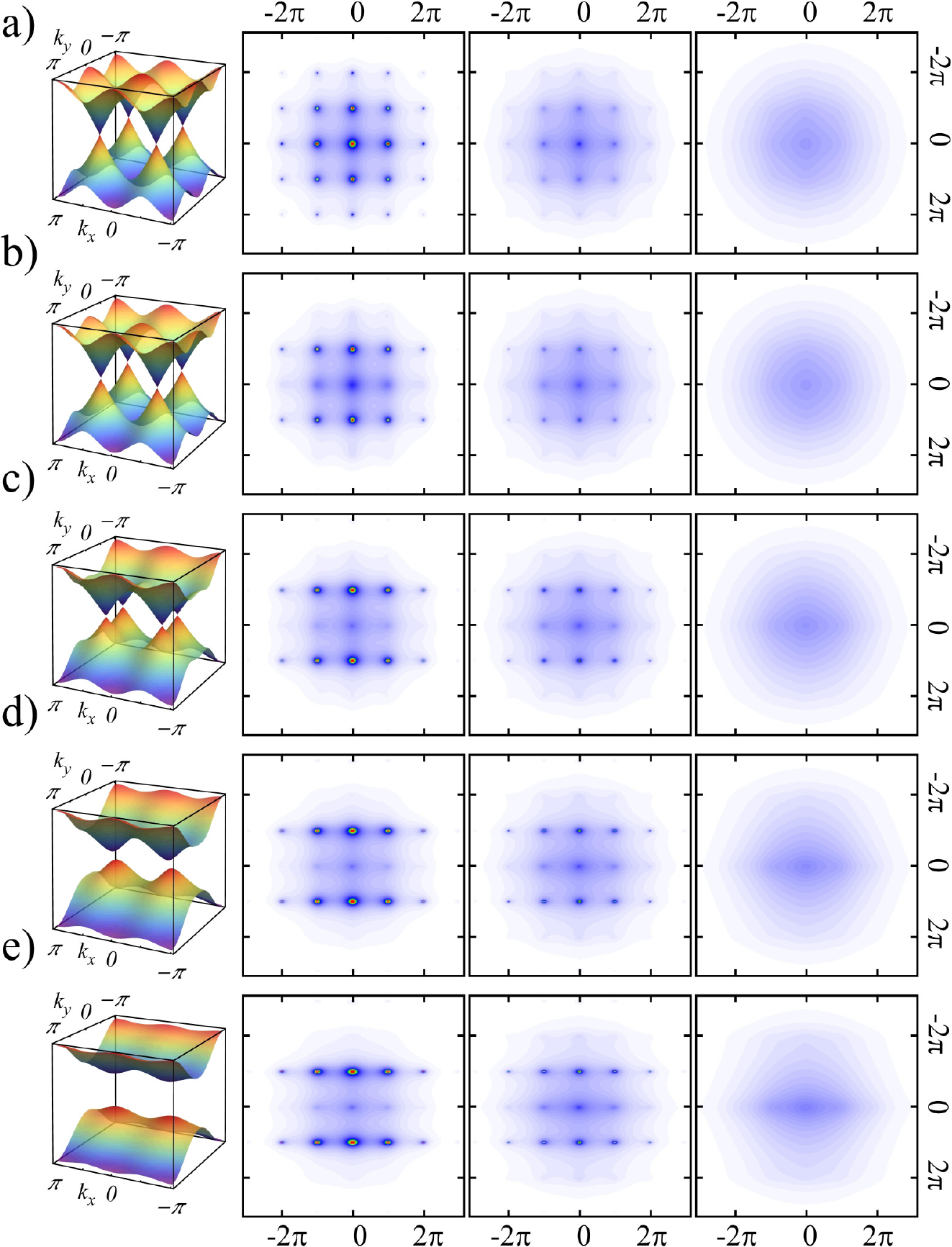}

\caption{(Color online) Dispersion relation and time-of-flight patterns for
atoms in optical lattice under uniform artificial magnetic field $f=1/2$
and additional staggering potential $\tilde{\Delta}$ driving particle
hopping for $\tilde{\Delta}$ equal to: a) 0, b) 0.1, c) 0.6, d) 1.2,
e) 2. The TOF images depict superfluid state (the first column, $V_{0}/E_{R}=5$),
system close to the phase transition (the second column, $V_{0}/E_{R}=7.5$)
and the Mott insulator (the last column, $V_{0}/E_{R}=10$).}

\label{fig:4}
\end{figure}

Application of the additional staggering potential $\Delta/2t\equiv\tilde{\Delta}$
that drives hopping of atoms between chosen lattice sites {[}see Sec.
II{]} is a natural extension of the system in the uniform magnetic
fields (see, Fig. \ref{fig:1}b). It allows to manipulate the Dirac
cones: change the distance between them in the $k$-space and annihilate
them when two of them merge \cite{delplace}. For $f=1/2$ (flux through
the elementary cell $\phi_{\mathrm{}}=\pi$) and $\tilde{\Delta}=0$,
the system is identical to described in Sec. \ref{sub:uniform_1p2}
and the resulting TOF patterns are presented in Fig. \ref{fig:4}a.
While the $\tilde{\Delta}$ is being increased, the Dirac points move
closer to each other (see, Fig. \ref{fig:4}b-c) and for $\tilde{\Delta}=1$
-- merge. For $\tilde{\Delta}>0$, the Dirac points annihilate and
the single-particle spectrum becomes gapped (Fig. \ref{fig:4}d-e).
However, as in the previous case of the uniform field, if both sub-bands
are populated, the condensation of bosons occurs around the bottom
of the lower band, thus Dirac cones have no effect on the TOF patterns.
On the other hand, increasing value of $\tilde{\Delta}$ strongly
enhances hopping along one direction leading to slow decline in weight
of $\mathbf{k}=0$ maximum, enlarging $\left(n\pi,\pm\pi\right)$
components (with $n\ne0$). This effect can be reversed by enlarging
interatomic interactions ($U/t$): near the superfluid -- Mott insulator
phase transition the mobility of atoms is naturally decreased. Once
again, this allows to observe the weak maxima in momentum distribution
of incoherent particles which are located around $\mathbf{k}=0$ regardless
of the position of the superfluid phase coherence peaks.

\begin{figure}
\includegraphics[scale=0.6]{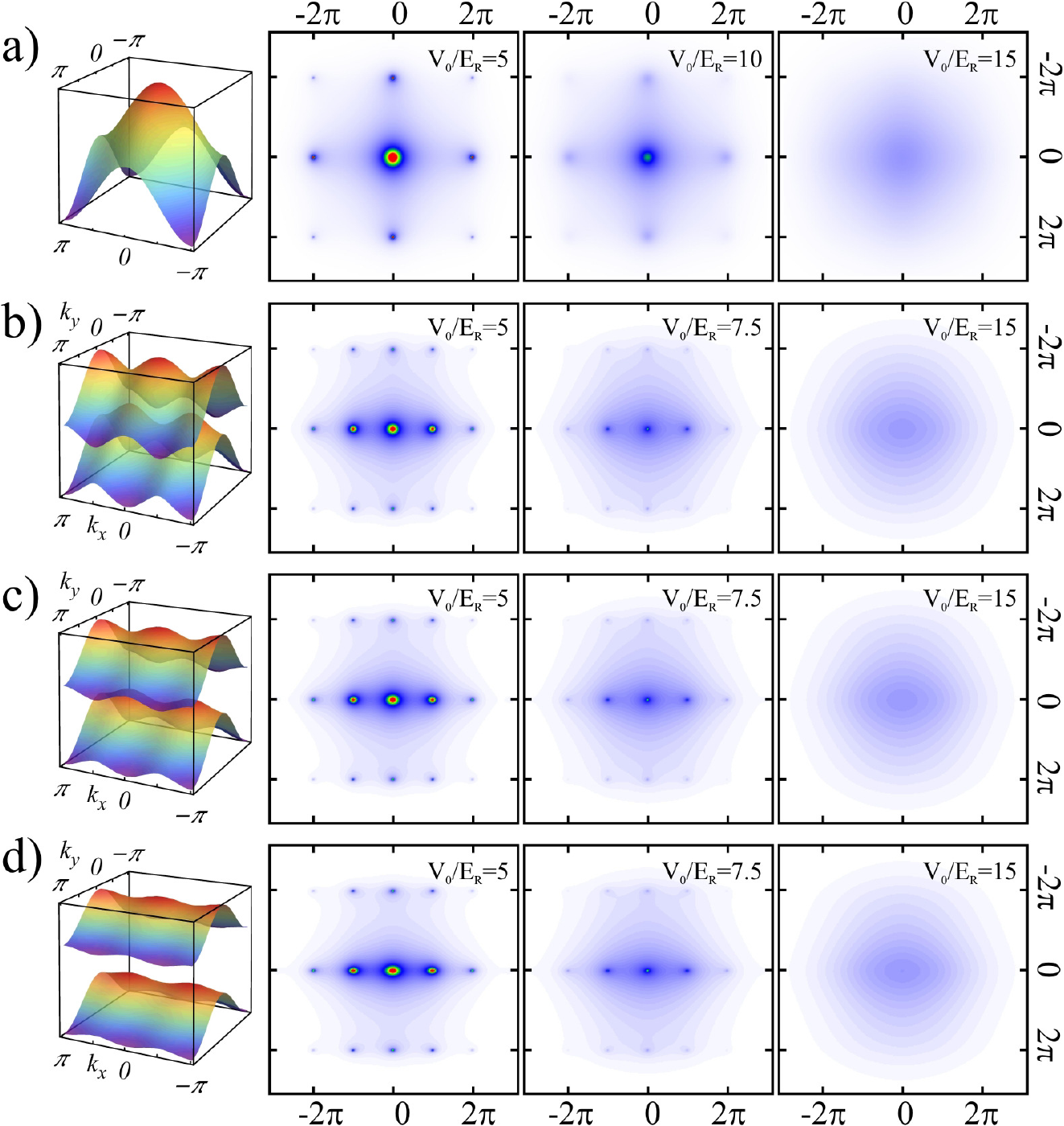}\caption{(Color online) Dispersion relation and time-of-flight patterns for
atoms in optical lattice under additional staggering potential $\tilde{\Delta}$
driving particle hopping for $\tilde{\Delta}$ equal to: a) 0, b)
0.6, c) 1.2, d) 2. The TOF images depict superfluid state (the first
column), system close to the phase transition (the second column)
and the Mott insulator (the last column). }
\label{fig:5}
\end{figure}

A slightly different behavior can be observed for a system without
artificial magnetic field, but with the staggering potential $\tilde{\Delta}$
(see, Fig. \ref{fig:5}). Since the flux is missing, the maxima occur
at $\mathbf{k}=0$, however their density in $k_{x}$ direction is
doubled due to increased size of an elementary cell (with the width
of the Brillouin zone halved). Increase of $\tilde{\Delta}$ leads
to smearing of the peaks in $k_{x}$ direction. The dispersion relations
are quite different: the Dirac cones are not present and the spectrum
is gapped for every value of $\tilde{\Delta}$. Also, a noticeable
difference occurs for $\tilde{\Delta}\approx0$: in the system with
$f=0$ the TOF images change discontinuously while going from non-zero
to zero value of $\tilde{\Delta}$. This results from the fact that
presence of the staggering potential breaks the translational symmetry
doubling the size of the elementary cell for every value of $\tilde{\Delta}$,
but not for $\tilde{\Delta}=0$ (see, differences between Figs. \ref{fig:5}a
and b-e). This is in contrast to the $f=1/2$ case, where the enlargement
of the elementary cell resulting from the presence of the $\pi$ flux
per plaquette and the staggering potential are the same. 

\subsection{Checkerboard staggered flux }

\begin{figure}
\includegraphics[scale=0.6]{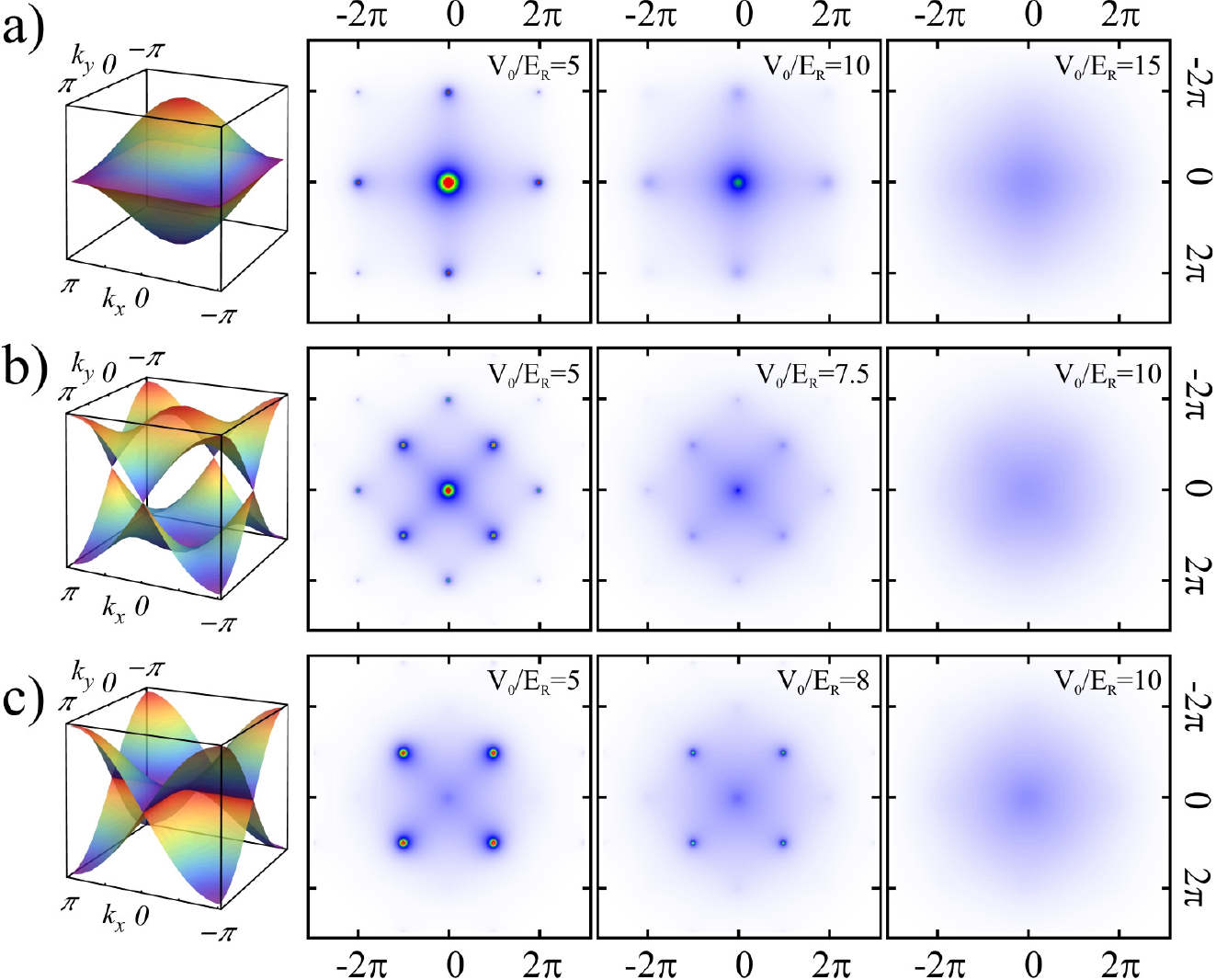}\caption{(Color online) Dispersion relation and time-of-flight patterns for
atoms in optical lattice in \emph{checkerboard} staggered artificial
magnetic field with flux $\phi$ per plaquette: a) $0$, b) $\pi$,
c) $2\pi$. The TOF images depict superfluid state (the first column),
system close to the phase transition (the second column) and the Mott
insulator (the last column).}

\label{fig:6}
\end{figure}

A time-independent lattice model with an artificial staggered magnetic
field that is used in the present work can effectively describe time-dependent
optical lattice with staggered particle current in the tight-binding
regime \cite{lim}. As a result, it is possible to describe group
of experiments that use temporal modification of the optical potential.
They allow reaching regimes, where anisotropic Dirac cones emerge
in the single-particle spectrum leading to two inequivalent conical
points in the energy band, which results in two distinct energy minima
that depend on the magnitude of the staggered magnetic flux $\phi$.
Consequently, it is possible to realize the artificial magnetic field
as presented in Fig. \ref{fig:1}c, where the flux is staggered and
arranged in checkerboard configuration. This method allows to reach
values of flux per plaquette ranging from $-2\pi$ to $2\pi$. The
resulting time-of-flight images are presented in Fig. \ref{fig:6}.
For small fluxes $\phi_{\mathrm{}}$ the effect of the magnetic field
is hardly noticeable, as the strong maximum at $\mathbf{k}=0$ is
visible. For $\phi=\pi$ (see Fig. \ref{fig:6}b) ) the energy minima
of the two conical points become equal, leading to peaks in momentum
distribution located in $\mathbf{k}=0$ and $\mathbf{k}=\left(\pm\pi,\pm\pi\right)$
points. Finally Fig. \ref{fig:6}c), for larger fluxes the non-zero
momentum state takes over reaching maximum intensity for $\phi=2\pi$.
It should be stressed that in a naive view the impact of the flux
$n2\pi$ ($n$ being integer) should be negligible. However, here
the $\phi=\pm2\pi$ flux leads to non-trivial superfluid phase with
non-zero momentum. This issue will be discussed in the following subsections.

\subsection{Uniaxially staggered flux\label{sub:Uniaxially-staggered-flux}}

\begin{figure}
\includegraphics[scale=0.6]{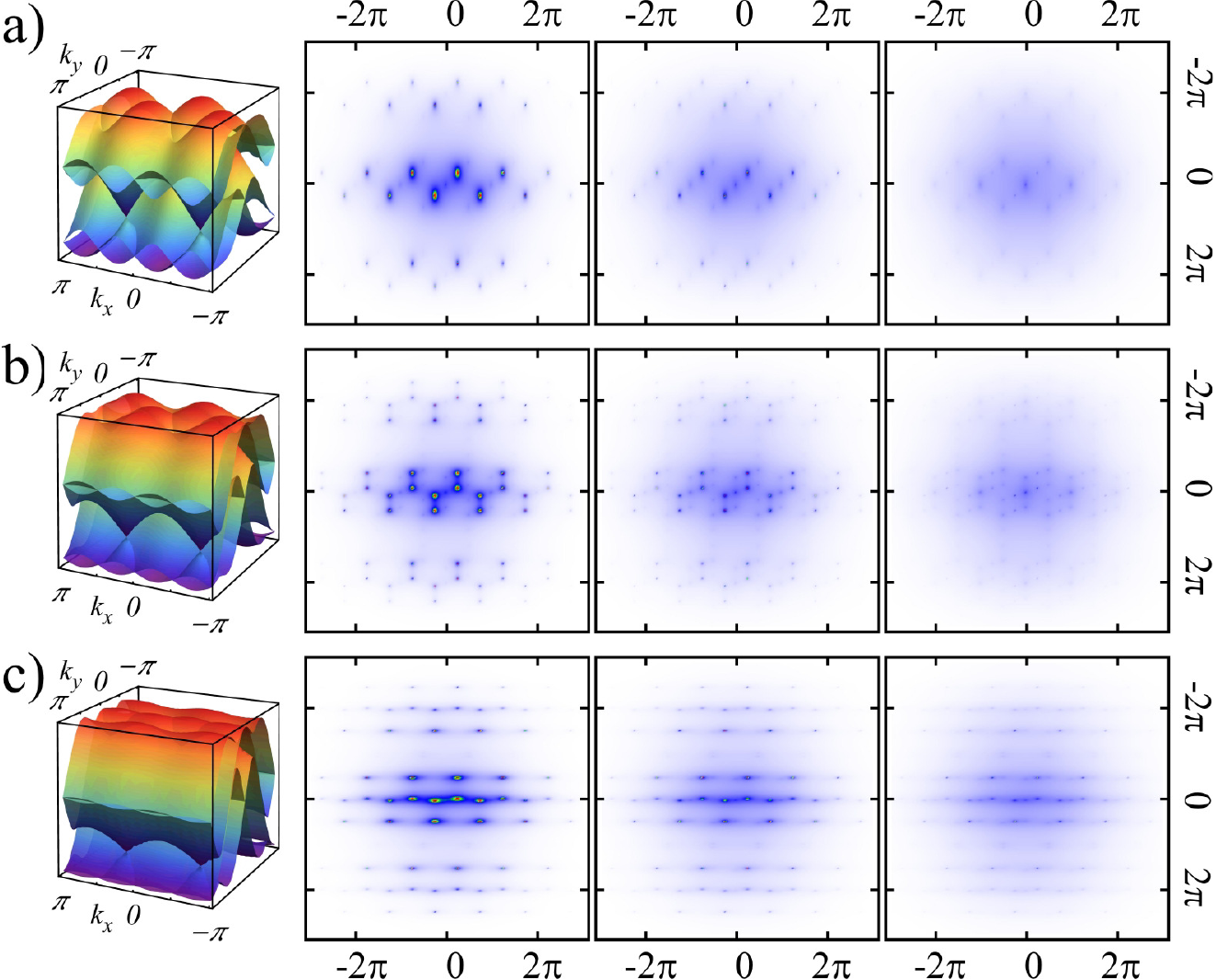}

\caption{(Color online) Dispersion relation and time-of-flight patterns for
atoms in optical lattice in uniaxially staggered artificial magnetic
field with flux per plaquette $\phi=\pi/2$ for hopping anisotropy
leading ratio of $t_{i+1,j}/t_{i,j+1}$ equal to : a) $1$, b) $2$,
c) $3$. The TOF images depict superfluid state (the first column),
system close to the phase transition (the second column) and the Mott
insulator (the last column).}
\label{fig:7}
\end{figure}

Photon-assisted tunneling in an optical superlattice generating large
tunable effective magnetic fields for ultra-cold atoms demonstrated
possibility of realization of the large tunable uniaxially staggered
field (where the spatial average of the flux is zero) \cite{aidelsburger}.
It was shown that the atomic sample relaxes to the minima of the magnetic
band structure, realizing an analogue of a frustrated classical spin
system. The obtained time-of-flight patterns for various system hopping
anisotropies \cite{aidelsburger,moller} agree well with the ones
calculated with the method presented in the current work (see, Fig.
\ref{fig:7}). Positions of the maxima of the momentum distribution
are correctly recreated as well as the decay of the envelope of the
TOF image substantiating choice of the module of the Wannier function
in Eq. (\ref{eq:wannier}). For isotropic system, the time-of-flight
patterns exhibit two minima located around the zero momentum at $\mathbf{k}=\pm\left(\pi/4,\pi/4\right)$.
While the anisotropy between raw hopping in the $x$ ($t_{i+1,j}$)
and $y$ ($t_{i,j+1}$) direction is introduced, for $t_{i+1,j}/t_{i,j+1}\le\sqrt{2}$,
the peaks split into pairs of peaks in agreement with changes of the
magnetic band structure. 

\subsection{Arbitrary gauge geometry}

\begin{figure}
\includegraphics[scale=0.8]{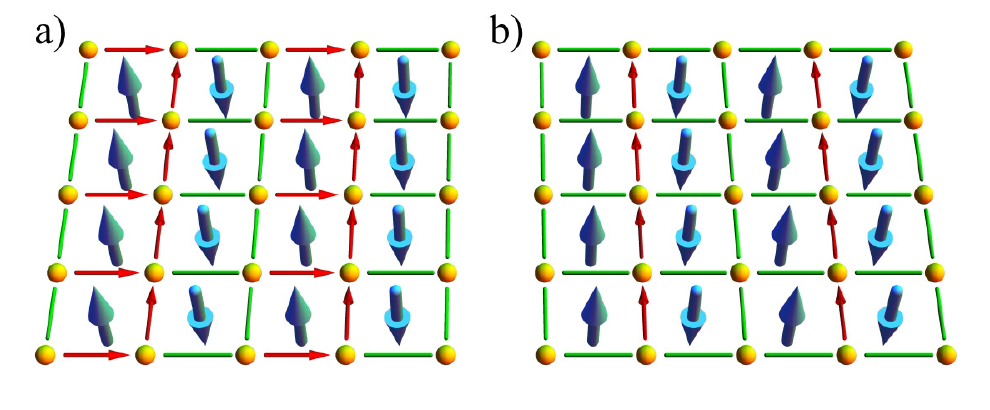}

\caption{(Color online) Comparison of the gauge used in Ref. \cite{aidelsburger}
and a simpler one resulting in the same configuration of the fluxes
(denoted by thick, blue arrows located inside the unit cell) $\pm\pi/2$,
while traveling around a plaquette. Thin arrows (red) denote direction
of hopping along which a particle phase changes by $\pi/2$ ($-\pi/2$
in an opposite direction), while tubes (green) represent regular hopping
with no phase change.}
\label{fig:8}
\end{figure}

\begin{figure}
\includegraphics[scale=0.6]{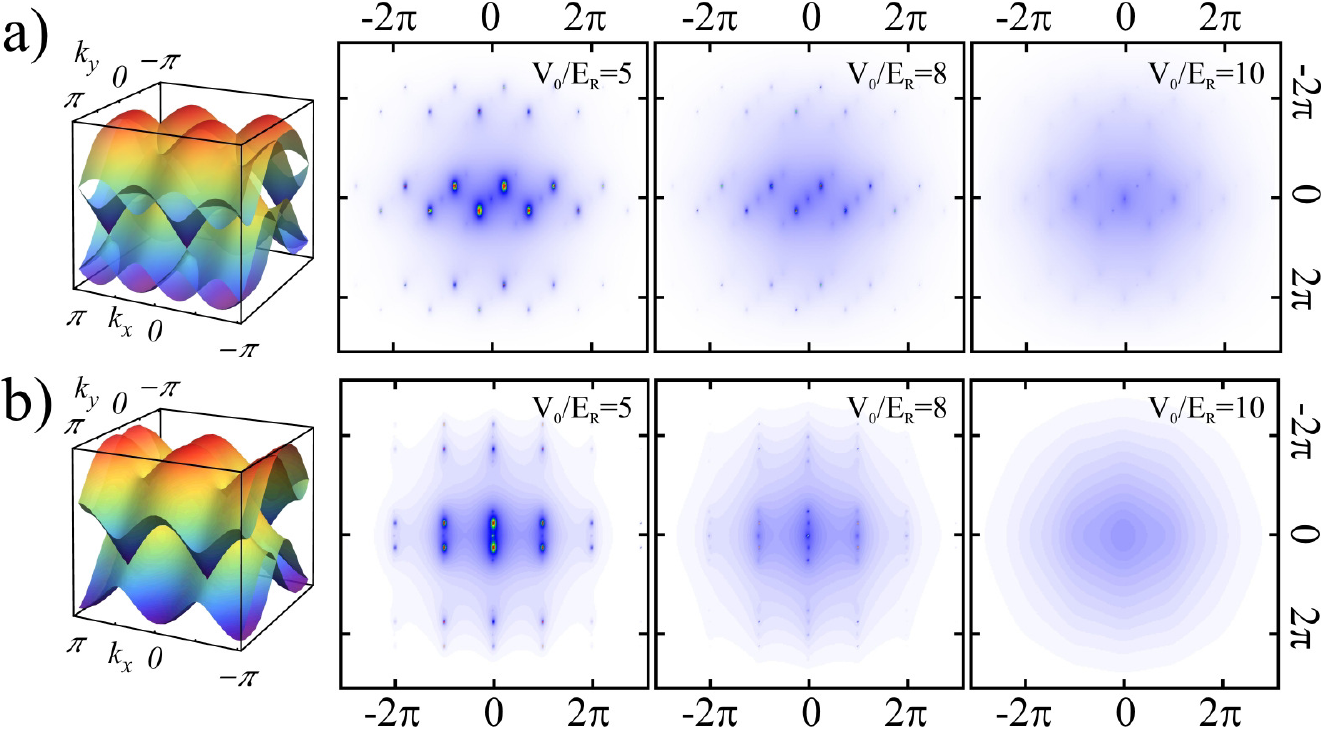}

\caption{(Color online) Comparison dispersion relation and time-of-flight patterns
for atoms in optical lattice in different gauges (see, Fig. \ref{fig:8}),
which lead to the same uniaxially staggered configuration of the magnetic
flux. }
\label{fig:9}
\end{figure}

\begin{figure}
\includegraphics[scale=0.6]{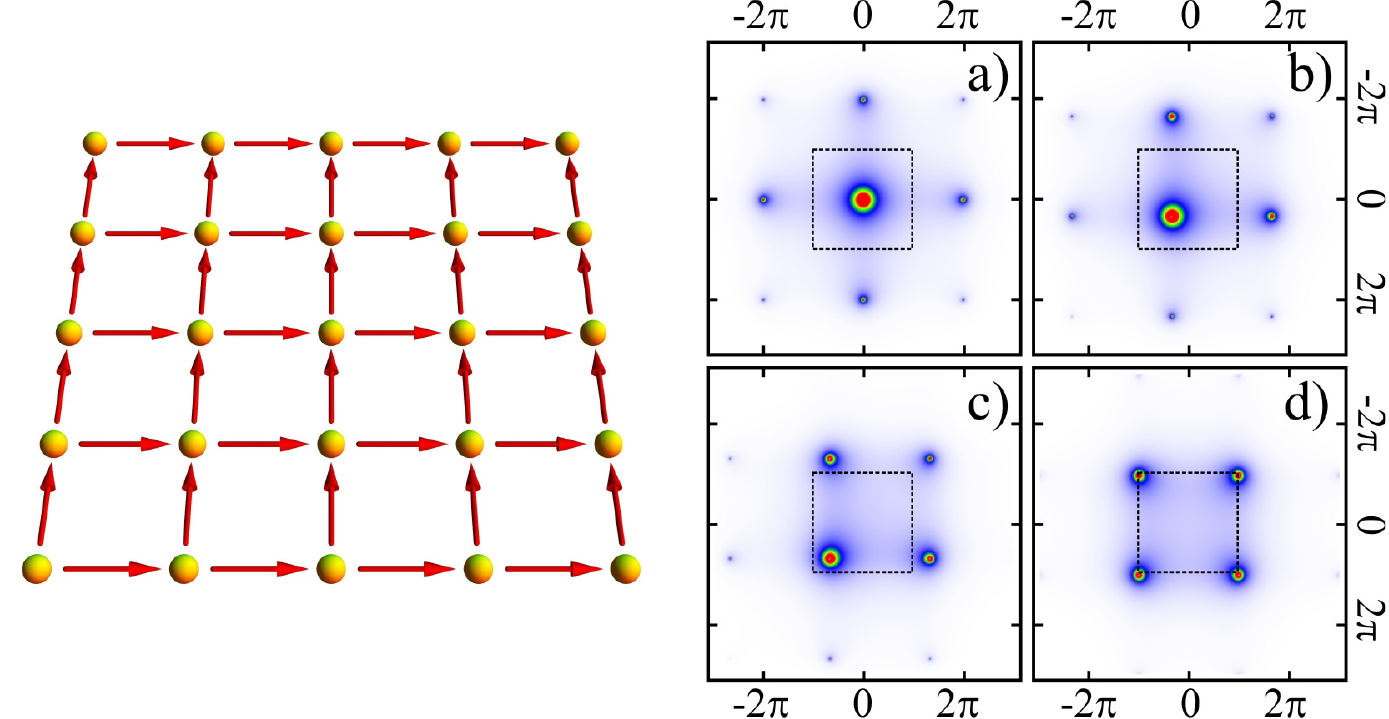}

\caption{(Color online) Gauge resulting in zero flux (red arrows denote direction
of hopping along which a particle phase changes by $\phi$) and resulting
time-of-flight patterns for atoms in optical lattice for $V_{0}/E_{R}=7.5$
and a) $\phi_{\mathrm{}}=0$, b), $\phi_{\mathrm{}}=\pi/3$, c) $\phi_{\mathrm{}}=2\pi/3$
and d) $\phi_{\mathrm{}}=\pi$. Dotted boxes mark the first Brillouin
zone.}
\label{fig:10}
\end{figure}

Although, the flux configuration in the case of the uniaxially staggered
flux (see, Sec. \ref{sub:Uniaxially-staggered-flux} and Fig. \ref{fig:1})
is pretty regular: uniaxially alternating values of $+\pi/2$ and
$-\pi/2$ every second plaquette, the structure of the TOF images
is complicated and strongly dependent on the lattice parameters like
hopping anisotropy. It results from complex gauge that was used in
the experiment in Ref. \cite{aidelsburger}. However, the same configuration
of the flux per plaquette can be obtained for much simpler gauge,
as presented in Fig. \ref{fig:8}b (the dispersion relation obtained
in the same manner as in Ref. \cite{aidelsburger}, however without
phase change along $k_{x}$ direction hopping). The comparison of
the resulting time-of-flight patterns is presented in Fig. \ref{fig:9}
and it is clearly visible that they are not identical. It can be seen
that in case of atoms moving in the tight-binding scheme in the optical
lattice the value of the flux being assigned to an elementary cell
\emph{does not determine} the momentum distribution of the particles.
Other than that, what is crucial is the \emph{change of the quantum
phase} that occurs at every bond that the particle travels along,
since all the jumps are separated acts rather than a continuous move.
This can lead to a non-intuitive situation of particles exhibiting
the influence of the vector potential field resulting of a non-zero
phase change on selected bonds, although the total phase change on
closed trajectory around an elementary cell is zero, which also means
that so is the effective flux per plaquette. Such situation is presented
in Fig. \ref{fig:10}: the configuration of the gauge results in change
of phase equal to zero when a particle travels around a closed loop.
However, the resulting time-of-flight patterns are still dependent
on the phase $\phi$ acquired during a single jump leading to dispersion
relation: 
\begin{equation}
t\left(\mathbf{k}\right)=t\left[\cos\left(k_{x}+\phi_{\mathrm{}}\right)+\cos\left(k_{y}+\phi_{\mathrm{}}\right)\right].
\end{equation}
For free condensate (without optical lattice) the superfluid velocity:
\begin{equation}
\mathbf{v}_{s}=\frac{\hbar}{m}\nabla\phi_{\mathrm{}}.
\end{equation}
The kinetic energy $E_{k}$ can be expanded around its minimum at
$\mathbf{k}_{\phi}=\left(-\phi_{\mathrm{}},-\phi_{\mathrm{}}\right)$,
which leads to $\mathbf{v}_{s}\sim\partial E_{k}/\partial k=\hbar\mathbf{k}_{\phi}/m$.
As a result, change of the particles momentum resulting from phase
acquired on a jump along a single bond is simply:
\begin{equation}
k\sim\nabla\phi_{\mathrm{}},
\end{equation}
which is consistent with the results in Fig. \ref{fig:10}. It is
worth to notice that using such simple gauge configuration one obtains
a finite momentum superfluid phase with the non-zero phase change
imposed on the condensate wave function but with the zero value of
the artificial magnetic field.


\section{Summary}

In conclusion, we have analyzed the correlations between strongly
interacting bosons confined in two-dimensional square lattice in the
presence of an artificial magnetic field using quantum rotor model
that is inherently combined with the Bogolyubov approach. The flexibility
of the method and its sensitivity to the spatial fluctuations allows
us to consider various geometries of the magnetic flux (uniform, checkerboard,
uniaxially staggered), which are expected to be realizable, or have
already been implemented in experimental settings. Furthermore, we
have calculated the time-of-flight patterns, which give information
about dynamics of condensed atoms surrendered to the artificial magnetic
field. We validate our approach by successfully recreating experimentally
observed TOF images and recovering the superfluid-Mott insulator phase
transition, which is driven by the interactions. Also, we show in
which conditions novel superfluid phases with non-zero momentum can
arise leading to observation of Dirac-like physics in optical lattices.
Furthermore, we deduce that the crucial element for a proper recreation
of the time-of flight patterns is not the flux configuration but rather
the change of the quantum phase that occurs at every bond described
by the dispersion relation and consequently the density of states.
Thus, the TOF images result directly from the chosen and experimentally
realized gauge. This is in clear contrast with solid state physics,
where attainable values of flux per elementary cell are very small
(flux $f=1/2$ would require fields of the order of $10^{2}-10^{3}$T).
As a result, the spatial change of the vector potential (gauge) is
very gradual and the phase acquired by particles on a single hop is
marginal. However, in optical lattices, the attainable values of the
flux are very high and the phase change on single bonds can be individually
controlled. This leads to the strong dependence of the time-of-flight
patterns and the atom dynamics on the specific gauge configuration
rather then the resulting magnetic flux. Therefore, in systems of
strongly interacting bosons, only the minima of the kinetic energy
(maxima of the dispersion relation in the $k$-space) determine the
superfluid properties of the ultra-cold atoms confined in optical
lattices. At the same time, the interactions between atoms change
the phase stiffness and the density of the condensate rather than
the dynamical properties of the coherent particles. 
\begin{acknowledgments}
T.P.P. would like to thank for the hospitality of I. Bloch's group
at MPQ in Garching and many fruitful discussions concerning experiments
with bosonic species, especially with M. Aidelsburger and the \emph{Bosons}.
T.P.P. would like to acknowledge partial funds from the Human Capital
Operational Programme, Grant No. UDA-POKL.04.01.01-00-133/09-00. We
would also thank M. Aidelsburger for insightful and helpful comments
regarding the manuscript. 
\end{acknowledgments}

\section{Appendix}

\subsection*{Quasiparticle spectrum and the density of states}

In the following, we present dispersion relations and the resulting
lattice densities of states for various flux geometries, which were
described in Sec. \ref{sec:Experimental-Gauges-of}.

\subsubsection{Staggered potential with uniform flux f=0 for two-dimensional square
lattice}

The dispersion relation contains two sub-bands \cite{delplace}

\begin{equation}
t_{0}\left(\boldsymbol{k}\right)=2t\left(\cos k_{y}\pm\sqrt{\cos^{2}k_{x}+\Delta^{2}}\right)
\end{equation}
and results from the single-particle Schr\"{o}dinger equation:
\begin{eqnarray}
E\psi_{m,n} & = & -t\psi_{m,n+1}-t\psi_{m,n-1}-t\psi_{m+1,n}\nonumber \\
 &  & -t\psi_{m-1,n}+\left(-1\right)^{m}\Delta\psi_{m,n},
\end{eqnarray}
which allows to deduce hopping elements $t_{mn}$ ($m,n$ number the
lattice sites in $x$ and $y$ directions, $E$ is the energy and
$\psi$ -- the wave function) between neighboring sites. The density
of states is given by a nonlinear convolution:
\begin{eqnarray}
\rho_{0}\left(E,\Delta\right) & = & \frac{1}{2\pi^{2}t}\int dx\rho_{1D}\left(x\right)\nonumber \\
 & \times & \rho_{1D}\left(\frac{E}{2t}\pm\sqrt{x^{2}+\Delta^{2}}\right),
\end{eqnarray}
where $\rho_{1D}\left(x\right)$ is a one-dimensional lattice density
of states: 
\begin{equation}
\rho_{1D}\left(E\right)=\frac{1}{2\pi^{2}}\frac{1}{\sqrt{1-\left(\frac{E}{2t}\right)^{2}}}
\end{equation}
resulting from $t\left(\boldsymbol{k}\right)=2t\cos k_{x}$ dispersion
relation.

\subsubsection{Staggered potential with uniform flux f=1/2 for the two-dimensional
square lattice}

The dispersion relation \cite{delplace}:
\begin{equation}
\left|t_{1/2}\left(\boldsymbol{k}\right)\right|=2t\sqrt{\cos^{2}k_{x}+\left(\cos k_{y}-\Delta\right)^{2}}
\end{equation}
 for $\Delta=0$ is equal to the uniform external magnetic field with
$f=1/2$. It results from the Schr\"{o}dinger equation:
\begin{eqnarray}
E\psi_{m,n} & = & -t\psi_{m,n+1}e^{i\pi m}-t\psi_{m,n-1}e^{-i\pi m}\nonumber \\
 & - & t\psi_{m+1,n}-t\psi_{m-1,n}+\left(-1\right)^{m}\Delta\psi_{m,n}.
\end{eqnarray}
The resulting DOS reads:

\begin{eqnarray}
\rho_{1/2}\left(E,\Delta\right) & = & \frac{\left|E\right|}{2\pi^{2}t}\int dx\frac{\rho_{1D}\left(x\right)}{\sqrt{\left(\frac{E}{2t}\right)^{2}-\left(x-\Delta\right)^{2}}}\nonumber \\
 & \times & \rho_{1D}\left(\sqrt{\left(\frac{E}{2t}\right)^{2}-\left(x-\Delta\right)^{2}}\right).
\end{eqnarray}

\subsubsection{Checkerboard staggered flux for the two-dimensional square lattice}

The dispersion relation for the flux $\phi$ is given by the formula
\cite{lim}:

\begin{eqnarray}
\left|t_{\phi}\left(\boldsymbol{k}\right)\right| & = & 2t\left[2\cos\left(\frac{\phi}{2}\right)\cos\left(\frac{k_{x}+k_{y}}{2}\right)\cos\left(\frac{k_{x}-k_{y}}{2}\right)\right.\nonumber \\
 & + & \left.\cos^{2}\left(\frac{k_{x}+k_{y}}{2}\right)+\cos^{2}\left(\frac{k_{x}-k_{y}}{2}\right)\right]^{1/2}
\end{eqnarray}
and results from the Schr\"{o}dinger equation:
\begin{eqnarray}
E\psi_{m,n} & = & -t\psi_{m,n+1}e^{i\left(-1\right)^{p}\frac{\phi}{4}}-t\psi_{m,n-1}e^{-i\left(-1\right)^{p}\frac{\phi}{4}}\nonumber \\
 & - & t\psi_{m+1,n}e^{-i\left(-1\right)^{p}\frac{\phi}{4}}-t\psi_{m-1,n}e^{i\left(-1\right)^{p}\frac{\phi}{4}}
\end{eqnarray}
with $p=m+n$. The DOS can be written in the form:
\begin{eqnarray}
\rho\left(E,\phi\right) & = & \frac{\left|E\right|}{2\pi^{2}t}\int\frac{dx}{\cos\frac{\phi_{\mathrm{}}}{2}+x}\rho_{1D}\left(x\right)\nonumber \\
 & \times & \rho_{1D}\left(\frac{E^{2}+x^{2}-1}{\cos\frac{\phi_{\mathrm{}}}{2}+x}-x\right)
\end{eqnarray}
where any value of flux $\phi$ is allowed.

\subsubsection{Uniaxially staggered flux}

The dispersion relation for gauge configuration used in experiments
presented in Ref. \cite{aidelsburger} consists of four sub-bands:
\begin{eqnarray}
t_{1,2}\left(\mathbf{k}\right) & = & \sin k_{x}-\eta\cos k_{y}\nonumber \\
 & \pm & \sqrt{\eta^{2}-2\sin2k_{x}+\eta^{2}\sin2k_{y}+2}\nonumber \\
t_{3,4}\left(\mathbf{k}\right) & = & t_{1,2}\left[\mathbf{k}-\left(\frac{\pi}{2},\frac{\pi}{2}\right)\right],\label{eq:dispaidels}
\end{eqnarray}
where $\eta$ is a hoping anisotropy ratio, while the dispersion used
in Fig. \ref{fig:8}b, reads:
\begin{eqnarray}
t_{1,2}\left(\mathbf{k}\right) & = & \cos k_{y}-\sin k_{y}-\sqrt{2\cos2k_{x}+\sin2k_{y}+3}\nonumber \\
t_{3,4}\left(\mathbf{k}\right) & = & t_{1,2}\left[\mathbf{k}-\left(\frac{\pi}{2},0\right)\right].
\end{eqnarray}
In both cases, analytical formulas for the lattice density of states
cannot be easily obtained. The details of calculation of the dispersion
in Eq. (\ref{eq:dispaidels}) can be found in Ref. \cite{aidelsburger}.


\begin{thebibliography}{10}
\bibitem{anderson}M. H. Anderson, J. R. Ensher, M. R. Matthews, C.
E. Wieman, and E. A. Cornell Science, \textbf{269}, 198 (1995).

\bibitem{greiner}M. Greiner, O. Mandel, T. W. H\"{a}nsch and I.
Bloch, Nature \textbf{415}, 39 (2002).

\bibitem{jaksch1}D. Jaksch, C. Bruder, J. I. Cirac, C. W. Gardiner,
and P. Zoller, Phys. Rev. Lett. \textbf{81}, 3108 (1998).

\bibitem{leggett}A. Leggett, \emph{Quantum liquids} (Oxford, New
York, 2006).

\bibitem{cooper}N. R. Cooper, Adv. Phys. \textbf{57}, 539 (2008).

\bibitem{jaksch}D. Jaksch and P. Zoller, New J. Phys. \textbf{5},
56 (2003).

\bibitem{gerbier}F. Gerbier and J. Dalibard, New J. Phys. \textbf{12},
033007 (2010). 

\bibitem{kolovsky}A. Kolovsky, Europhys. Lett. \textbf{93}, 20003
(2011).

\bibitem{lin}Y.-J. Lin, R. L. Compton, K. Jiménez-García, J. V. Porto
and I. B. Spielman, Nature, \textbf{462}, 628 (2009).

\bibitem{aidelsburger}M. Aidelsburger, M. Atala, S. Nascimbène, S.
Trotzky, Y.-A. Chen, and I. Bloch, Phys. Rev. Lett. \textbf{107},
255301 (2011); M. Aidelsburger, M. Atala, S. Nascimbène, S. Trotzky,
Y.-A. Chen, I. Bloch, arXiv:1212.2911.

\bibitem{osterloh}K. Osterloh, M. Baig, L. Santos, P. Zoller, and
M. Lewenstein, Phys. Rev. Lett. \textbf{95}, 010403 (2005).

\bibitem{hauke}P. Hauke, O. Tieleman, A. Celi, C. \"{O}lschl\"{a}ger,
J. Simonet, J. Struck, M. Weinberg, P. Windpassinger, K. Sengstock,
M. Lewenstein and A. Eckardt, Phys. Rev. Lett. \textbf{109}, 145301
(2012).

\bibitem{harper}P. Harper, Proc. Phys. Soc. London Sect. A \textbf{68},
874 (1955). 

\bibitem{hofstadter}D. Hofstadter, Phys. Rev. B \textbf{14}, 2239
(1976).

\bibitem{zaleski}T. A. Zaleski and T. K. Kope\'c, Phys. Rev. A \textbf{84},
053613 (2011).

\bibitem{zaleski1}T. A. Zaleski, Phys. Rev. A \textbf{85}, 043611
(2012); T. A. Zaleski, J. Phys. B: At. Mol. Opt. Phys. \textbf{45},
145303 (2012).

\bibitem{polak}T. P. Polak and T. K. Kope\'c, Phys. Rev. \textbf{A}
79, 063629 (2009).

\bibitem{sinha}S. Sinha and K. Sengupta, Eur. Phys. Lett. \textbf{93},
30005 (2011).

\bibitem{polak1}T. P. Polak and T. K. Kope\'c, Phys. Rev. B \textbf{76},
094503 (2007).

\bibitem{polak2}T. P. Polak and T. K. Kope\'c, J. Phys. B: At. Mol.
Opt. Phys. \textbf{42,} 095302 (2009).

\bibitem{polakzaleski1}T. A. Zaleski, T. P. Polak, Phys. Rev. A \textbf{83},
023607 (2011); T. P. Polak, T. A. Zaleski, Acta. Phys. Pol. A \textbf{121},
1312 (2012).

\bibitem{capogrosso-sansone}B. Capogrosso-Sansone, \c{S}. G\"{u}ne\c{s}
S\"{o}yler, Nikolay Prokof'ev and B. Svistunov, Phys. Rev. A \textbf{77},
015602 (2008).

\bibitem{teichmann}N. Teichmann, D. Hinrichs, M. Holthaus, and A.
Eckardt, Phys. Rev. B \textbf{79}, 100503 (2009). 

\bibitem{oktel}M. \"{O}. Oktel, M. Ni\c{t}\u{a}, and B. Tanatar,
Phys. Rev. B \textbf{75}, 045133 (2007).

\bibitem{niemeyer}M. Niemeyer, J. K. Freericks, and H. Monien, Phys.
Rev. B \textbf{60}, 2357 (1999). 

\bibitem{blackie}P. B. Blakie and C. W. Clark, J. Phys. B: At. Mol.
Opt. Phys. \textbf{37}, 1391, (2004).

\bibitem{vojta}T. Vojta, Phys. Rev. \textbf{B} 53, 710 (1996).

\bibitem{kato}Y. Kato, Q. Zhou, N. Kawashima and N. Trivedi, Nature
Physics \textbf{4}, 617 (2008).

\bibitem{powell}S. Powell, R. Barnett, R. Sensarma, and S. Das Sarma,
Phys. Rev. Lett. \textbf{104}, 255303 (2010); S. Powell, R. Barnett,
R. Sensarma, and S. Das Sarma, Phys. Rev. A \textbf{83}, 013612 (2011).

\bibitem{pollet}L. Pollet, N. V. Prokof'ev, and B. V. Svistunov,
Phys. Rev. Lett. \textbf{104}, 245705 (2010).

\bibitem{aidelsburger1}M. Aidelsburger and \emph{Bosons, }private
communication.

\bibitem{denschlag}J. Hecker Denschlag, J. E. Simsarian, H. H\"{a}ffner,
C. McKenzie, A. Browaeys, D. Cho, K. Helmerson, S. L. Rolston and
W. D. Phillips, J. Phys. B: At. Mol. Opt. Phys. \textbf{35}, 3095
(2002).

\bibitem{fallani}L. Fallani, F. S. Cataliotti, J. Catani, C. Fort,
M. Modugno, M. Zawada, and M. Inguscio, Phys. Rev. Lett. \textbf{91},
240405 (2003).

\bibitem{delplace}P. Delplace and G. Montambaux, Phys. Rev. B \textbf{82},
035438 (2010).

\bibitem{moller}G. M\"{o}ller and N. R. Cooper, Phys. Rev. A \textbf{82},
063625 (2010). 

\bibitem{lim}Lih-King Lim, C. Morais Smith, A. Hemmerich, Phys. Rev.
Lett. \textbf{100}, 130402 (2008); Lih-King Lim, A. Hemmerich, and
C. Morais Smith, Phys. Rev. A \textbf{81}, 023404 (2010).\end{thebibliography}
\end{document}